\documentclass[11pt]{article}
\usepackage[cp1251]{inputenc}
\input{epsf}

\def\be{\begin{equation}}
\def\ee{\end{equation}}
\def\ba{\begin{eqnarray}}
\def\ea{\end{eqnarray}}

\def\12{{1\over 2}}

\def\msun{M_\odot}

\def\ltsima{$\; \buildrel < \over \sim \;$}
\def\simlt{\lower.5ex\hbox{\ltsima}}
\def\gtsima{$\; \buildrel > \over \sim \;$}
\def\simgt{\lower.5ex\hbox{\gtsima}}

\begin{document}

\title{\bf The Minimum Stellar Mass in Early Galaxies}
\author{E.~O.~Vasiliev$^{1,2}$, Yu.~A.~Shchekinov$^1$\thanks{yus@phys.rsu.ru}\\
\it $^1$Rostov State University, Rostov-on-Don, Russia\\
\it $^2$Physics Research Institute, Rostov State University,
Rostov-on-Don, Russia}

\date{}

\maketitle

\begin{abstract}
The conditions for the fragmentation of the baryonic component during merging
of dark matter halos in the early Universe are studied. We assume that the
baryonic component undergoes a shock compression. The characteristic masses
of protostellar molecular clouds and the minimum masses of protostars
formed in these clouds decrease with increasing halo mass. This may
indicate that the initial stellar mass function in more massive galaxies was
shifted towards lower masses during the initial stages
of their formation. This would result in an increase of the number of stars per
unit halo mass, i.e., the efficiency of star formation.
\end{abstract}

\maketitle



\section{INTRODUCTION}

\noindent
In the hierarchical scenario of structure formation
in the Universe, gravitationally bound objects -- dark
matter halos, in which stars can form -- first appear
at redshifts $z\sim 20$ and have masses $M \sim 10^6~\msun$,
including the dark and baryonic components [1–3].
These halos form due to multiple collisions and 
merging of similar objects of smaller mass. The
dynamics of the early stages of these processes are
determined by the gravitational instability of the dark
component, which consists of cool (non-relativistic)
particles. Density perturbations of practically any
wavelength can develop in the dark component, since
the minimum critical scale of the perturbations -- the
Jeans mass, which depends on the velocity dispersion
in a collisionless gas -- is negligible compared with
the mass of the halo (see, for example, [4]).
The spatial scale of the perturbations can be constrained from below
by a truncation of the perturbation spectrum. The
amplitude of the density perturbations on small scales
is known to decrease with wavelength as
a power law $\propto \lambda^{-3}$ [5, 6]. Thus, the formation
of structure includes relatively developed shortwavelength
motions modulated by slowly growing long-wavelength flows [7].

Under these conditions, a nearly one-dimensional
motion develops, in which perturbations are compressed
mainly along a single direction, giving rise
to a disk-like configuration [8, 9]. Baryons confined by
the gravitation field of the dark matter and involved
in this one-dimensional flow undergo shock compression
and are heated to the temperature $T_f \sim m_p
v_c^2/2k$, where $v_c$ is the velocity of the inward gas
flow; as a result, a dense layer or disk is formed.
The compressed baryonic layer is efficiently cooled by
radiative processes by neutral hydrogen atoms
or molecules. The formation of molecular hydrogen
is substantially enhanced behind shock fronts, so
that the cooling of the gas may basically be governed
by the H$_2$ molecules. When the temperature decreases
to 200~K, deuterium efficiently transforms
into HD, which results in an even higher thermal energy
loss rate, providing favorable conditions for the
fragmentation of the compressed gas [10–17].
Thus, during the formation of galaxies, converging flows
result in an emergence of dense baryonic
condensations in which stellar objects can subsequently
form. This concept was recently discussed
in [16, 17]. During the formation of massive protogalaxies,
dense baryonic layers are maintained over
longer times, so that the gas is cooled substantially
and higher densities are reached. Therefore, one may expect
that during the formation of more massive
galaxies, baryonic fragments of lower mass can become
gravitationally unstable, and less massive stars can
be formed, which increases the number of stars
per unit mass of the halo -- in other words, the star formation
efficiency. Here, we study this possibility.

Section 2 discusses the model adopted to describe
the thermal evolution of baryons behind shock
fronts. Section 3 presents the results, and Section
4 contains our conclusions and final discussion.
Our calculations were based on the  $\Lambda$CDM
model for the Universe: $(\Omega_0,\Omega_{\Lambda},
\Omega_m,\Omega_b,h ) = (1.0,\ 0.71,\ 0.29,\ 0.047,\
0.72 )$, with the assumed deuterium abundance $n[D]/n =
2.6\cdot 10^{-5}$ [18].


\section{HALO VIRIALIZATION AND THE THERMO-CHEMICAL EVOLUTION OF GAS}

\noindent
In the linear stage of its growth, the evolution
of a dark-matter density perturbation $\delta_{dm}(t)$ can be
described by the equation (see e.g. [6]):
\be
\ddot\delta_{dm}+2H\dot\delta_{dm} = {3\over 2}H^2(\Omega_b\delta_{b} +
\Omega_{dm}\delta_{dm}),
\label{dm}
\ee
where $H$ is the Hubble constant, $\delta_{b}$ is the perturbation
of the baryon density, and $\Omega_{dm} = \Omega_{m} - \Omega_{b}$. As
seen from this equation, there is no limiting
minimum scale in dark matter. In reality, the Jeans
mass in collisionless dark matter depends on the
velocity dispersion; however, this value turns out to
be negligibly small compared with the mass of the
halo. As was already noted, if a density perturbation
in dark matter is considerably overcritical, then
its compression will proceed more rapidly along one
direction [8, 9]. The dissipative baryonic component
follows the dark matter potential, and forms a disk-like
configuration in the symmetry plane. The growth of the
perturbation is essentially a collision between flows
of dark matter and baryons, and results in the formation
of a shock front in the baryonic component. Subsequently,
the collisionless dark
matter becomes virialized via violent relaxation [19],
and the baryonic component undergoes shock compression
in the colliding flows, forming a dense radiatively
cooling layer. Starting from some time, the
increased density of the gas may exceed the density
of the dark matter, and the dynamics of the gas will
be determined by its own parameters. Therefore, we can
neglect the influence of the dark matter and consider
only the baryonic component. This is valid at least
within the characteristic collision time between
regions, $t_d \sim D/v$.

When gas flows collide, a thin dense gas layer
and two diverging shock fronts around the contact
region are initially formed; behind the shock fronts,
the gas is heated to the temperature $T_f \sim m_p {v_c}^2/2k$,
where $m_p$ is the proton mass and $v_c$ the collisional
velocity. The compression and formation of the dense
layer lasts approximately $t \sim t_{d} = D/v_c$. In
the transverse direction, the gas is not restrained
by dynamical pressure and can expand freely on a
time scale $t_{\perp} \sim D/c_s$, where $c_s$ is the sound speed.
Thus, the ratio of the characteristic time scales is
$t_d/t_{\perp} \simeq 1/{\cal M}$, where  ${\cal M}$ is the Mach number, and
we have for a collision between baryonic flows with
velocities $v_c>c_s$ the condition $t_d<t_{\perp}$. Therefore,
for supersonic collisions, we can neglect the transverse
motion in the layer when $t<t_d$ and solve a
one-dimensional problem. It was shown via two-dimensional
modeling that, indeed, only an insignificant
amount of the mass is lost to transverse outflow
during a collision of gas clouds [20].
The velocity of the gas motions during the formation
of a dark halo with mass $M$ is
\begin{equation}
\label{vel}
v_c = \sqrt{3}\sigma,
\end{equation}
where
\begin{equation}
\label{velvir}
  \sigma(M) = \sqrt{GM^{2/3}(3\pi^3 \Omega_m \rho_0)^{1/3}(1+z)},
\end{equation}
$\sigma$ is the one-dimensional velocity dispersion and $\rho_0$ is
the present mean total density of the Universe. It is
clear that the collisional velocity and gas temperature
behind the shock fronts will be higher when more
massive halos are formed.

The thermal evolution of baryons behind a shock
front can be described by a system of ordinary differential
equations written for a single Lagrangian element of the fluid:
\be
\label{el}
\dot x = \beta x n (1 - x - 2f) - k_1 n x^2,
\ee
\be
\label{mole}
\dot f = k_m n(1-x-2f)x,
\ee
\be
\label{molehd}
\dot g = k_{D1} f x n d_c - n x (k_{D1} f + k_{D2}) g,
\ee
\be
\label{tempe}
\dot T = {2\over 3}{\dot n \over n}T + \Sigma\Lambda_i.
\ee
where $x=n(\rm{e})/n$, $f=n(\rm{H}_2)/n$, $g=n(\rm{HD})/n$ are
the relative concentrations of electrons, H$_2$, and HD,
$d_c = n(\rm{D})/n$ is the cosmological abundance of deuterium,
$k_i$ are the rates of reactions [3, 21], $\beta$ is the rate
of collisional ionization of hydrogen [14], $\Lambda_i$ is the rate
of cooling and heating due to Compton interactions
with photons of the cosmic microwave background
radiation (CMB), emission in atomic and molecular
hydrogen lines [22], and in HD lines [21, 23, 24].
Initially, the gas density behind the shock front is
$n=4n_0$, where $n_0$ is the density before the shock
front. Further, we will assume that each element of
the gas behind the shock front evolves isobarically,
and that the density is described by the equation
\be
\label{iso}
n={p\over \mu kT},
\ee
where $\mu = \rho / n m_p$.

In the $\Lambda$CDM scenario, the spectrum of perturbations
in the dark matter is rather steep ($n=-3$),
which implies that halos with masses of $10^4-10^8~\msun$
condense out over times shorter than the corresponding Hubble
time [25]. A certain time after forming,
the halos become virialized. In addition, since they
are involved in large-scale motions, they merge with
each other, forming halos of higher mass. Both the virialization
and the merging occur on comparable times. For this
reason, two limiting possibilities can be suggested for
the initial conditions for the halo formation. In the
first, halos of smaller mass (subhalos) collide and
soon reach the virial state at larger redshifts. In the
second, virialization and merging of subhalos occur
at the same time at a given redshift. In both cases,
the process can be presented as a one-dimensional
compression. The two cases differ in the initial characteristics
of the matter in the flow; more precisely, in
the density and the velocity of the collision. In the first
model, a halo is formed due to collisions of subhalos,
and the parameters of the matter correspond
to those in the objects that have been virialized by
the beginning of the formation of the larger halo ($z_{ta}$),
i.e., by the time it separates from the cosmological
expansion. The matter density can be taken to be
$\rho \simeq 18\pi^2 \rho_0(z_{ta})$ [25], and the relative concentrations
of e, H$_2$, and HD to their values inside the subhalos.
In the second model, subhalos are virialized
and collide at the same redshift, and the initial density
of the matter is $\rho \simeq 18\pi^2 \rho_0(z_{vir})$, where $\rho_0(z)$ is the
background density of matter at the redshift $z$; the
relative concentrations of electrons, H$_2$, and HD are
assumed to be equal to their values inside the halo.
In both models, the collision results in the formation
of shock fronts, behind which gas is rapidly heated
to the temperature $T_f = m_p v_c^2/3k$, which is taken as
the initial temperature of an element of gas behind
the front. Since the initial conditions in the models
considered differ only in their densities, the results will
be qualitatively similar. Therefore, we will describe
only the first model, but present graphs for both. Calculations
for halos with the masses corresponding to
3$\sigma$ perturbations were carried out only for the second
model.


\section{RESULTS}

\subsection{Thermal and Chemical Evolution of Baryons}

Let us consider the chemical and thermal evolution
of the gas behind the shock front using the
example of halos formed at the redshift $z = 15$.
Figure 1 presents the concentrations of electrons, H$_2$,
and HD and the temperatures reached in the gas
behind the shock front in the end of one compression time
during the formation of a halo with mass $M$. It is clearly seen
that, starting from a certain value of halo mass, the relative concentration
of H$_2$ molecules exceeds $5\cdot 10^{-4}$ and the
temperature of the gas decreases substantially ($T\sim 500$~K).
This corresponds to the minimum H$_2$ concentration
needed in order for the gas to cool rapidly, in one comoving Hubble
time [3]. In more massive
halos, the H$_2$ concentration continues to grow, and
can reach $10^{-2}$ for halos with $M_h \simgt 4\cdot 10^7~\msun$.
Radiative losses in H$_2$ lines cool the gas to $T\simlt 200~$K,
and all the deuterium becomes rapidly bound in HD
molecules due to the effects of chemical fractionation.
Due to strong cooling in HD lines, the temperature
drops to several tens of Kelvin, which is close to the
temperature of the CMB radiation at this redshift
$2.73 (1 + z)$. HD molecules provide efficient heat
exchange between the CMB and baryons, due to
the fact that they absorb background photons
and subsequently transfer the excitation energy to
the gas through collisions [24, 26, 27]. Under these
conditions, the Jeans mass behind the shock front
(dashed curves in lower panels of Fig. 1) decreases
and, starting from some value, becomes substantially
lower than the baryonic mass of the forming halo
($\Omega_b M_h/\Omega_m$, solid curve). For example, for a halo with
$M = 10^7~\msun$, the Jeans mass is only $M_J \simeq 10^3~\msun$.

Let us now consider the instability of the cool
compressed layer and possible formation of dense
baryonic condensations in it. We will assume that the
gas in the layer is gravitationally unstable and can
fragment provided the critical perturbation length $\lambda_m$
is shorter than the initial size of the cloud $D$, and
the corresponding time $t_m$ is shorter than the
compression time [28, 29]:
\ba
\label{inscriterion}
\lambda_m/D \leq 1,~~~ t_m/t_d \leq 1.
\ea
The vertical line in Fig. 1 indicates the minimum
halo mass for which the criterion (\ref{inscriterion})
is fulfilled during
its formation. Thus, the gas behind the shock front
originating during the formation of a halo with mass
$M\ge 10^7\msun$ at a redshift $z \simeq 15$ is unstable by the
end of the compression phase; accordingly, the first
baryonic objects with masses roughly equal to the
Jeans mass $M_J\leq 10^3\msun$ can form inside the
halo. Since the temperature of the gas in the unstable
fragments is lower than $200$~K, the abundance
of HD also increases substantially, thereby determining
the subsequent thermal evolution of the gas behind
the shock fronts forming during the formation of the
first halos.

Figure 2 shows the interrelation between the redshift
and the minimum halo mass for which the layer behind
the shock front becomes during its formationunstable according to
the criterion (\ref{inscriterion}). For comparison,
the mass of 3$\sigma$ perturbations is displayed, as well as
the minimum mass for which baryons in already virialized halos can cool and
form gravitationally bound objects [3]. For instance,
a gravitationally unstable gas
layer is formed during the virialization behind the
shock front in a halo with mass $M_{min} \simeq 5\cdot
10^6~\msun$ at $z = 20$. Note that this derived minimum halo mass is comparable
to that obtained in [3], however, the temperature
of the gas is substantially lower in our model.

Baryonic objects with masses approximately equal
to the Jeans mass can form inside this unstable
layer. Figure 3 displays the dependence of the Jeans
mass behind the shock front on the time at which
the halo forms; the dashed line corresponds to the
minimum halo mass needed in order for the layer
behind the shock front to be unstable according to (\ref{inscriterion}),
and the solid curve to the mass of 3$\sigma$ perturbations.
It is seen that at the given redshift, the Jeans
mass in the layer formed behind the shock front decreases
with increasing halo mass: the conditions are
favorable for forming fragments with lower masses
in more massive halos. For example, in the formation
of a halo with mass $10^7~\msun$ (which corresponds
to the minimum halo mass) at redshift $z = 15$, the
layer behind the shock front becomes unstable, and
the baryonic objects formed there can have masses
of $M_b\sim 3\cdot 10^3~\msun$, while $M_b\sim 300~\msun$ for a halo
with mass $10^8~\msun$. Fragments with masses as low as
$M_b\sim 100~\msun$ will be unstable in a halo with mass
$3\cdot 10^9~\msun$ formed at $z = 10$. The possible masses
of gravitationally unstable baryonic condensations lie
within the above limits which widen with decreasing $z$.

During the formation of a massive halo, the gas
behind the shock front cools and is compressed
more intensely, due to the more efficient formation
of H$_2$ molecules behind more strong shock fronts.
Figure 4 presents the dependence of the gas density
on the mass of the halo and the redshift at which it
begins to form. The variation of the slope of the $n(M)$
curves with increasing $M$ reflects the fact that, by
the end of the compression phase, the temperature
behind the shock front reaches the temperature of the
CMB, due to the efficient formation of HD molecules
and energy exchange between the gas and CMB
radiation via the absorption of background photons by
HD molecules followed by collisional deexcitation.
The final density at a given $z$ depends only on
the initial temperature of the gas behind the shock
front, which, in turn, is a function of the halo mass:
$n = n_0(T_f/T_{CMB})$.

\subsection{The Minimum Mass of Baryonic Objects}

Baryonic fragments with masses $10^2-10^4~\msun$  (Fig. 3)
formed behind the shock front in an unstable
layer are optically transparent to radiation in H$_2$
and HD lines. Compression of the fragments occurs
in an isothermal regime at temperature close to
$2.73(1 + z)$~K, as is mainly ensured by the HD
molecules.

In the isothermal regime, a characteristic self-similar
density profile is formed, $\rho \sim r^{-2}$  [30, 31], the
optical depth in the HD lines increases in the central
regions, and the Jeans mass gradually decreases. The
critical point is the formation of an optically thick
($\tau \geq 1$) core region with a mass equal to or exceeding
the Jeans mass. In the central regions when $\tau \sim 1$,
the gas density is high, ($\sim 10^9-10^{10}$~cm~$^{-3}$), and
all the hydrogen is transformed into the molecular
phase via three-body reactions. However, due to
the low temperature of the gas, the contribution of
H$_2$ molecules to the cooling in optically transparent
regions -- and, therefore, their impact on the dynamics
of the compression -- is unimportant. The
formation of an opaque core is fully determined
by the HD molecules in the gas. After the formation
of an opaque, gravitationally unstable core, the
compression regime changes from isothermal to
adiabatic. Figure 5 presents the dependences of the
density, optical depth, temperature, and Jeans mass
inside the cloud when this state has been reached
by $z \simeq 15.3$ for a fragment with mass $M = 500~\msun$
formed as a result of instability in the layer during
the formation of a halo with $M_h = 2\cdot 10^7~\msun$ at
$z_v=17$. It is clear that the temperature through
the cloud is almost
constant and close to that of the CMB. The formed opaque core has a density of
$\geq 10^{10}$~cm$^{-3}$ and a mass of $\sim 0.15~\msun$.

Figure 6 presents the dependence of the mass of
the opaque core on the mass of the halo for the limiting
masses: the minimum mass $M_{\rm min}$ needed for the
layer behind the shock fronts to be unstable according
to (\ref{inscriterion}) for both models of the initial conditions, and the
mass corresponding to 3$\sigma$ perturbations, $M_{3\sigma}$. The
numbers in Fig. 6 denote the redshift at which the
halo is virialized. It is seen that the mass of the opaque
core decreases as the mass of the halo increases: the
mass of the core for a halo mass of $M_h \simeq 2\cdot 10^7~\msun$ at $z=17$
is $M(\tau \geq 1)\sim 0.15~\msun$, while the mass
of the core is only $M(\tau \geq 1)\sim 0.06~\msun$ for a halo mass
$M_h \simeq 10^9~\msun$ at $z=12$.

Thus, the formation of dark halos results in the formation
of strong shock fronts in the baryonic component,
substantial cooling, and an increase in the gas
density. The cold postshock gas layer is unstable against
formation of baryonic condensations in which protostellar
cores can in turn be formed. The higher the mass
of the initial halo, the lower the characteristic mass of
the gravitationally unstable fragment -- a possible
protostellar cluster or an opaque protostellar core.

\subsection{Thermal Instability}

The development of thermal instability becomes
possible in a radiatively cooling gas  under certain
cooling regimes [32]. In a non-steady-state case,
when radiative losses are not balanced by heating, the
condition for this is
\be
\label{thermins}
{d{\rm ln}\Lambda \over d{\rm ln}T} < 2,
\ee
where $\Lambda$ is the effective radiative-loss function (see,
e.g., [33]). Figure 7 shows the temperature
dependence of the function $d{\rm ln}\Lambda / d{\rm ln}T$ in the whole
range of temperatures encompassed by the gas cooling behind
the shock front. It is obvious that thermal
instability can develop always whenever the cooling
is determined by H$_2$ and HD molecules, i.e., when
70~K$<T<8000$~K. To order of magnitude, the characteristic
size of the region of instability coincides
with the size of the region behind the shock front
where the gas is cooled from $T=T_f$. Therefore, it is
obvious that the forming through thermal instability
condensations will facilitate
the subsequent gravitation fragmentation of the compressed
layer.

\section{DISCUSSION AND CONCLUSIONS}

\noindent
Collisions of baryonic flows during the formation
of the first protogalaxies are accompanied with an intense
cooling of the gas, which promotes fragmentation
of the gas onto condensations with characteristic
masses close to the masses of the present day
molecular clouds. The subsequent cooling and compression
of such condensations may result 
star formation. The higher the mass of the galaxy,
the more intense the gas cooling and the smaller
the mass of the molecular cloud and the minimum
mass of stars formed through subsequent
fragmentation. Thus, in this picture, we expect that
the initial mass function of stars and initial luminosity
function of stellar clusters will be shifted towards
smaller values in more massive galaxies.

The formed low-mass fragments cool rapidly, due
to high abundances of H$_2$ and HD molecules.
Starting from certain time, a protostellar cloud
converges to isothermal compression phase, since
the gas temperature is maintained near the CMB temperature
due to an efficient heat exchange between the CMB radiation and
baryons, which absorb background photons and then
transfer the excitation energy to the gas in collisional
processes [24, 26, 27]. Further, as the density in the
central regions of the fragment increases, an opaque
core with a mass of ($10^{-1}-10^{-2}$)$\msun$ is formed.
This core evolves then into a hydrostatic
protostellar core with a lower mass of the order
of $10^{-3}\msun$ [30, 31], onto which matter gradually
accretes, with an accretion rate determining the final
mass of the star [34]. In this process, practically all
gravitational energy of the accreted gas
is radiated in H$_2$ and HD lines. A possibility to observe
H$_2$ line emission was discussed in [35–
38]; it was found, however, that it is extremely difficult to detect this
emission with either currently operating or planned
telescopes [38]. For example, if $\sim 2000$ protostellar
objects are formed behind the shock front during
the virialization of a halo with mass $3\cdot 10^7\msun$
(which corresponds to approximately 10\% of the
mass of baryons being transformed into protostellar
fragments), the luminosity of this cluster in the
H$_2$ 2.34$\mu$m line will be $\sim 10^{38}$~erg/s. With a spectral
resolution of $R \sim 1000$, this corresponds to a flux
from the object of $\sim 10^{-2}$~$\mu$Jy at redshift $z = 15$,
while the sensitivity of the next planned SAFIR space
telescope\footnote{http://safir.jpl.nasa.gov}) is 1~$\mu$Jy.

\section*{ACKNOWLEDGMENTS}

The authors thank the referee for useful comments.


\newpage
\begin{figure}
\epsfxsize=15cm
\epsfbox{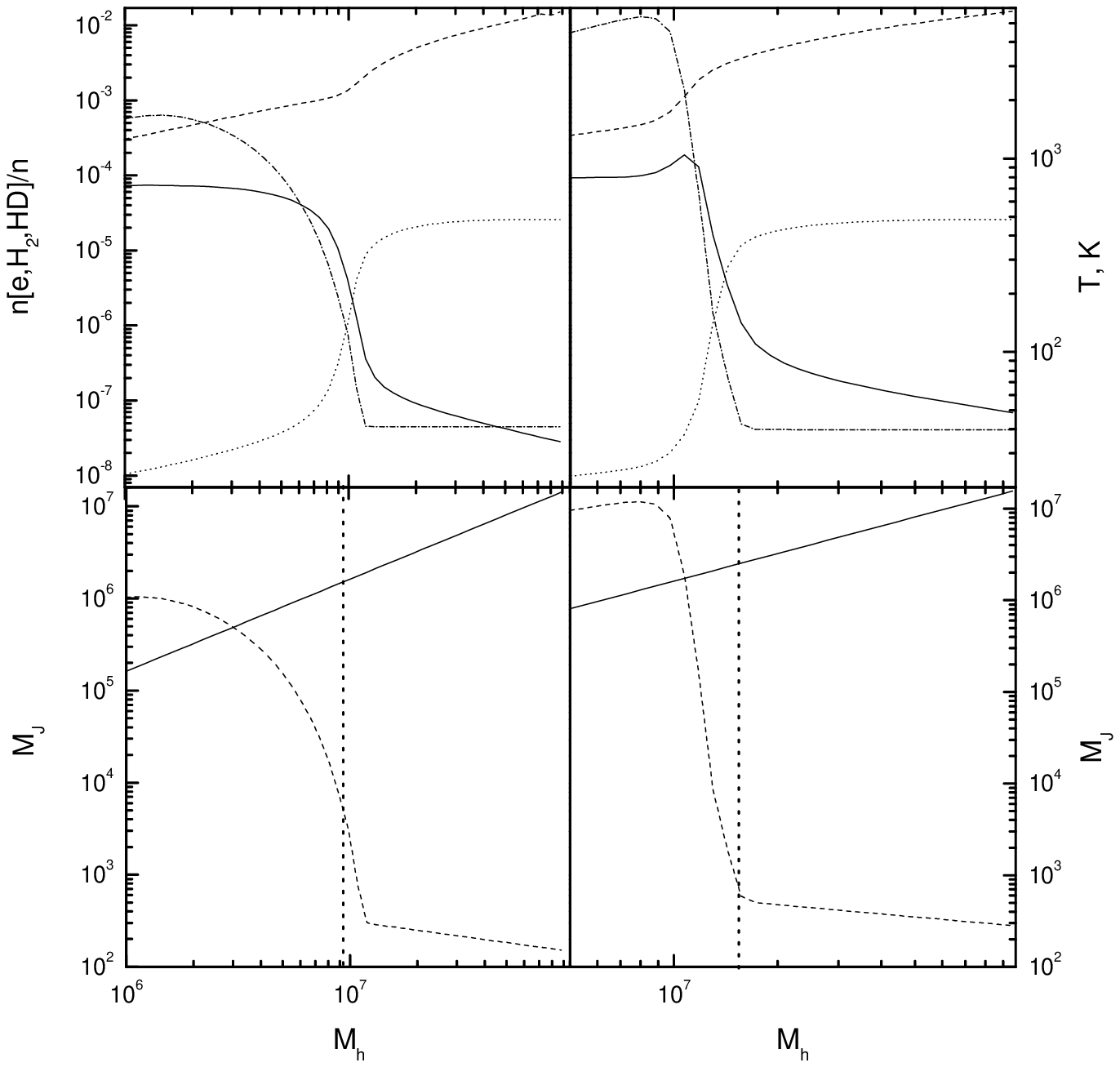}
\caption{Top: concentrations of electrons (solid curve), H$_2$ (dashed curve),
and HD (dotted curve), and the temperature (dot–dashed curve) reached
in the gas during virialization vs the halo mass $M$ at $z = 15$.
Bottom: total baryonic mass in the
halo (solid curve) and the Jeans mass in the layer behind the shock front.
The left and right panels correspond to the first and
second models for the initial conditions (see the text for more details).}
\label{z15}
\end{figure}

\begin{figure}
\epsfxsize=12cm
\epsfbox{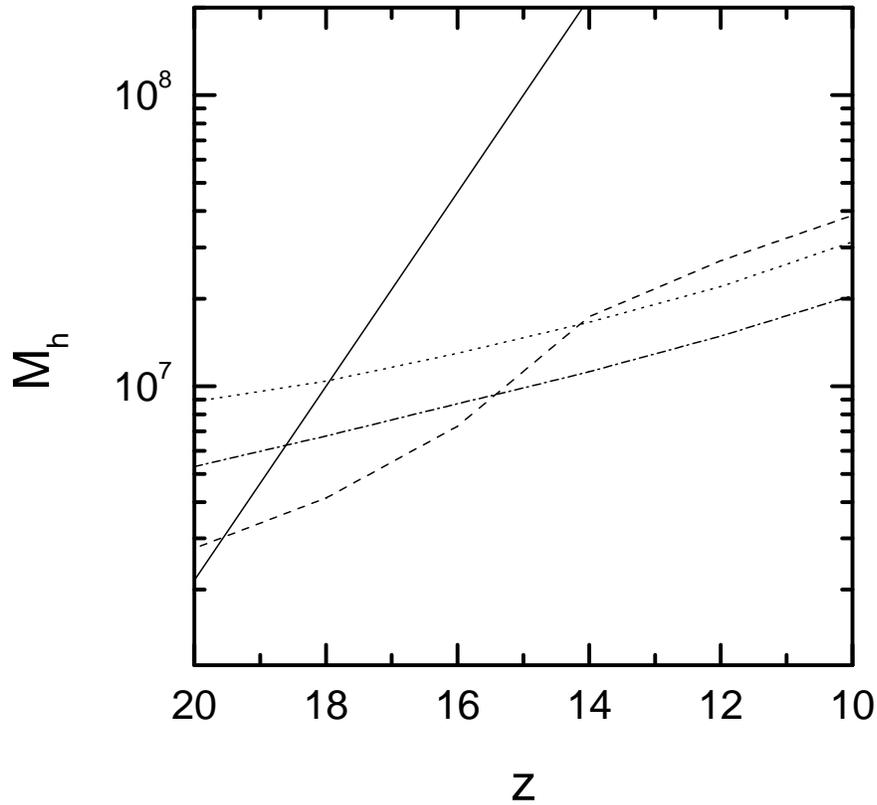}
\caption{The minimum halo mass for which the gas layer
behind the shock front becomes unstable according to the
criterion (\ref{inscriterion}) during the formation of
the halo for two models of halo formation
(dot–dashed and dotted curves). The
solid line indicates the mass of 3$\sigma$ perturbations, and the
dashed line the minimum mass for which baryons can
cool and form gravitationally bound objects in virialized halos [3].}
\label{minmass}
\end{figure}

\begin{figure}
\epsfxsize=12cm
\epsfbox{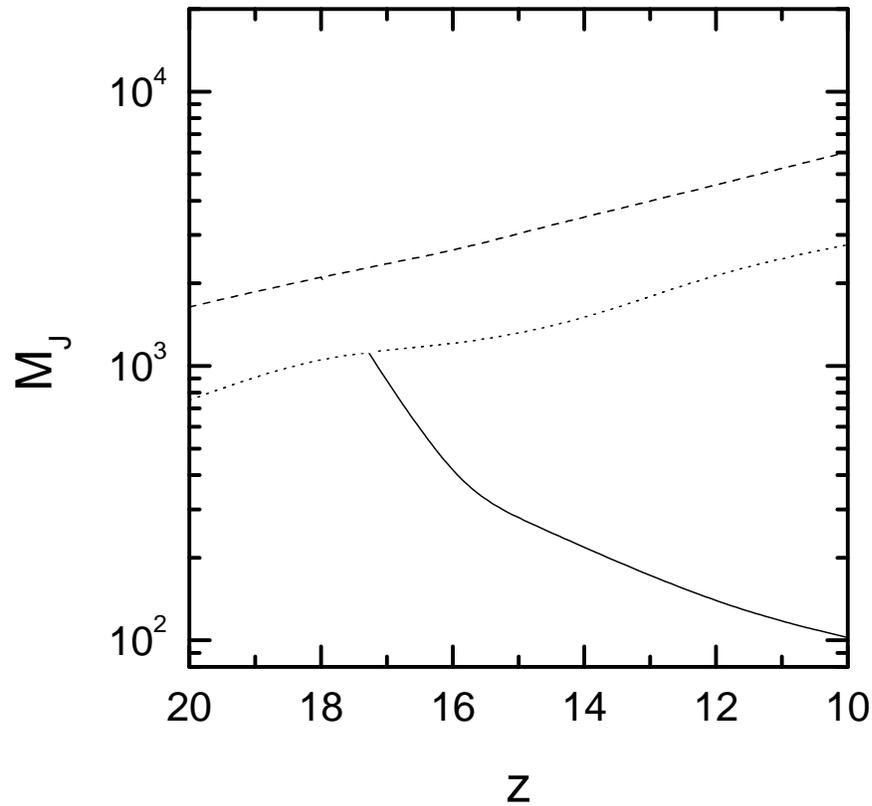}
\caption{Dependence of the Jeans mass behind the shock
front on the time when the halo forms. The dashed and
dotted lines correspond to the minimum halo masses
needed in order for the layer behind the shock front to
be unstable according to criterion (\ref{inscriterion}) for the first and
second models, respectively. The solid line corresponds to
the mass of the 3$\sigma$ perturbations.}
\label{mjeans}
\end{figure}

\begin{figure}
\epsfxsize=12cm
\epsfbox{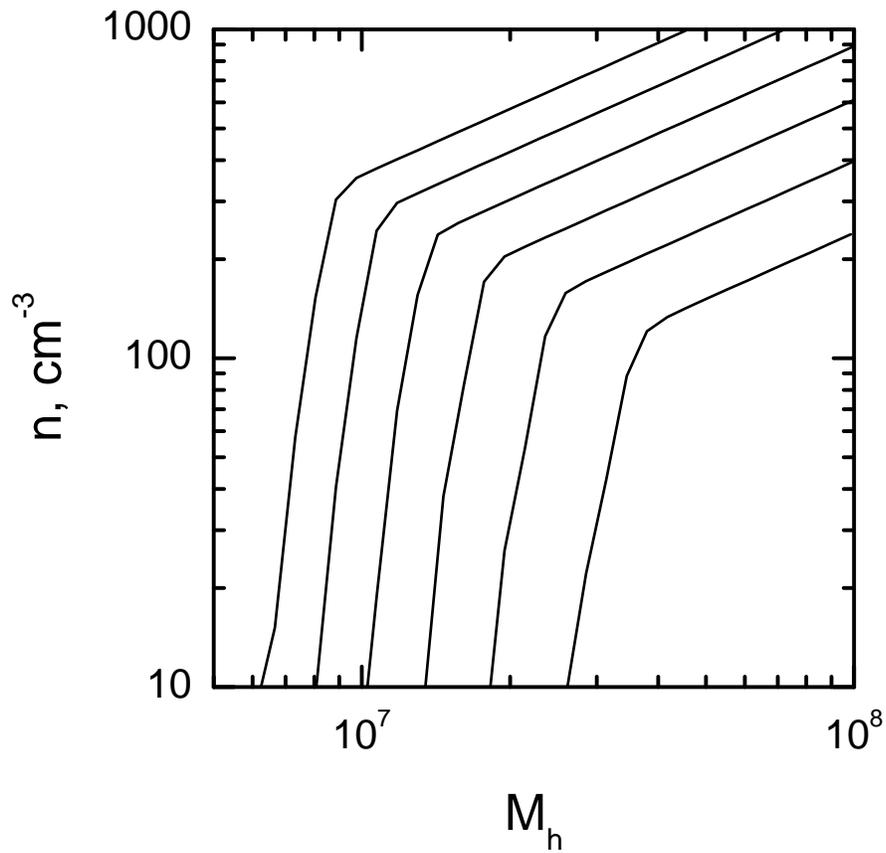}
\caption{Dependence of the final density of the gas behind
the shock fronts on the halo mass and the initial redshift at
which the halo forms for $z = 20, 18, 16, 14, 12, 10$
(left to right).}
\label{den}
\end{figure}

\begin{figure}
\epsfxsize=15cm
\epsfbox{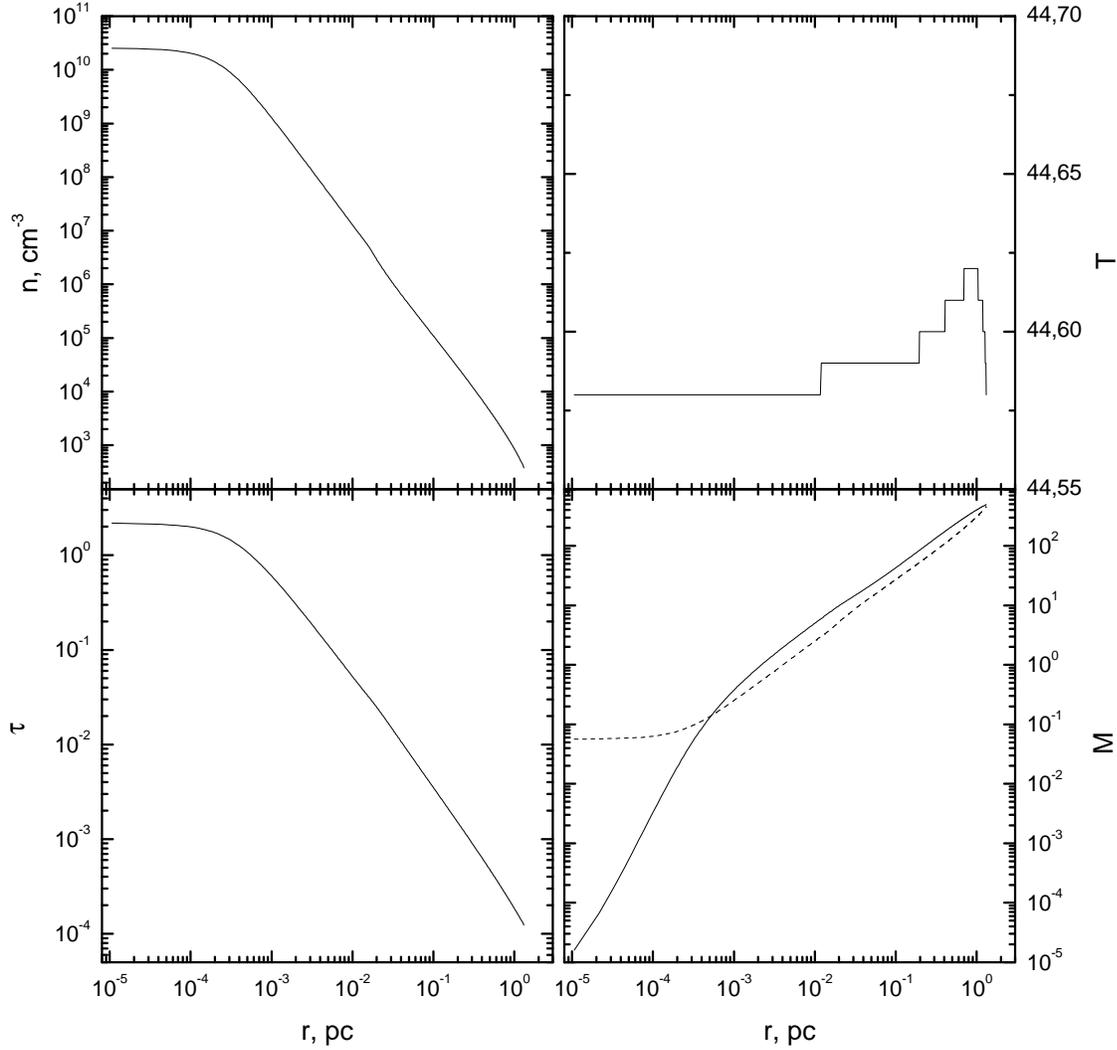}
\caption{Density, optical depth, temperature, and Jeans mass inside
a cloud at the time of formation of an opaque gravitationally
unstable core ($z \simeq 15.3$) for a fragment with mass $M = 500~\msun$
formed due to instability of the layer during the formation of a
halo with $M_h = 2\cdot 10^7~\msun$ at $z_v=17$.
The dashed line in the lower right panel indicates
Jeans mass, the solid line depicts
the mass inside the corresponding radius.}
\label{1d}
\end{figure}

\begin{figure}
\epsfxsize=12cm
\epsfbox{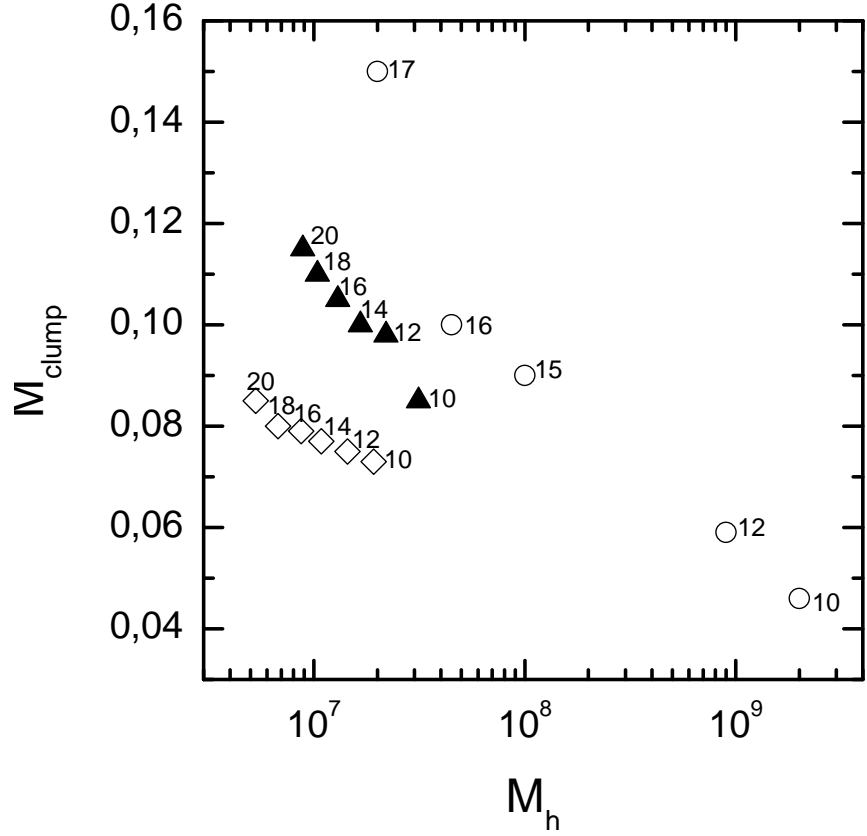}
\caption{Dependence of the mass of the opaque core on the
mass of the halo. The case of the minimum mass $M_{\rm min}$
needed for the layer behind the shock front to be unstable
according to (\ref{inscriterion}) for the first and second models for the
initial conditions are indicated by open diamonds and
filled triangles, respectively. The case of the mass corresponding to
3$\sigma$ perturbations, $M_{3\sigma}$, is indicated by open
circles. The numbers near the symbols denote the redshift
of the halo virialization.}
\label{mclump}
\end{figure}

\begin{figure}
\epsfxsize=14cm
\epsfbox{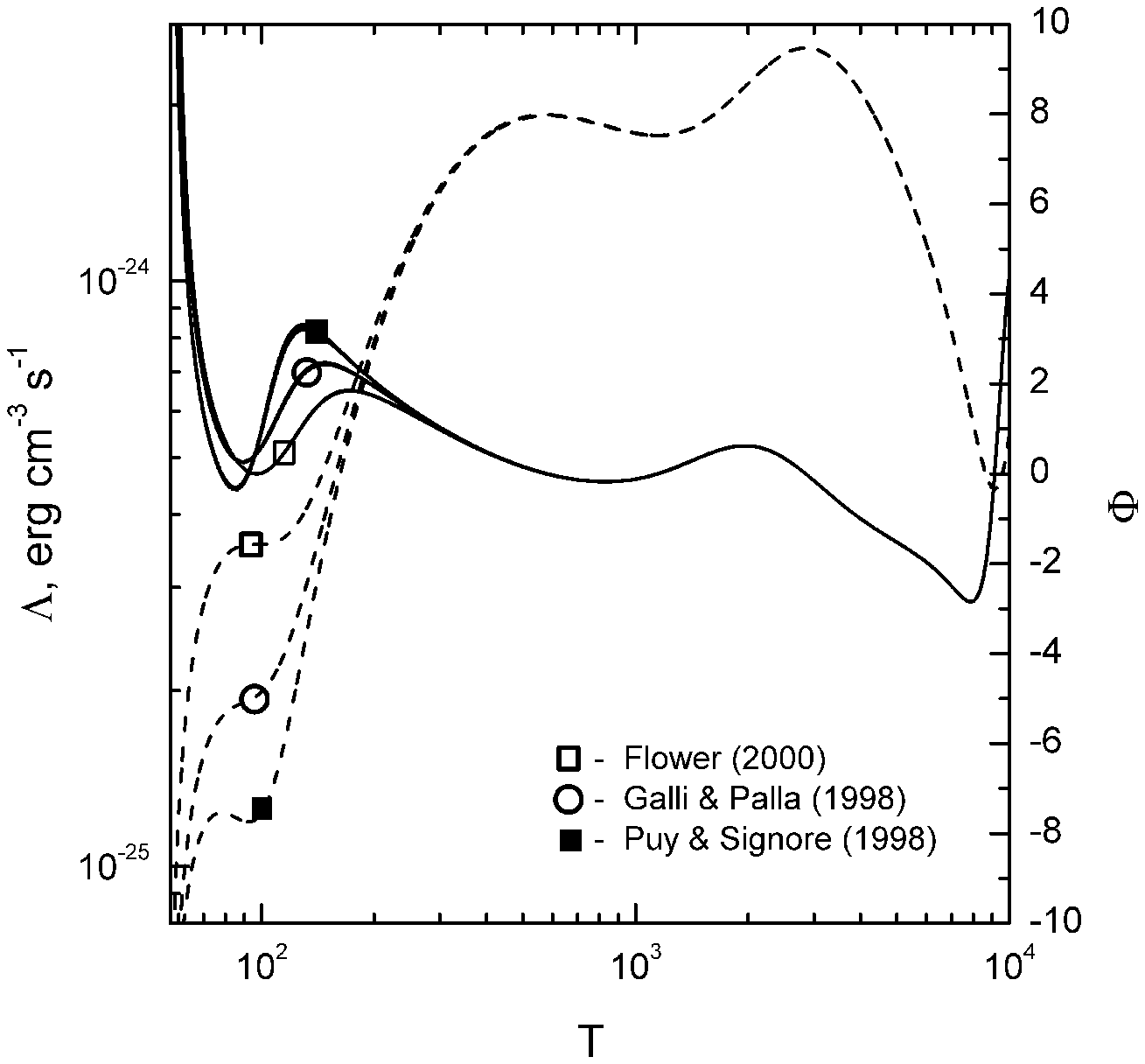}
\caption{Temperature dependence of the cooling rate and
its logarithmic derivative $\Phi = d{\rm ln}\Lambda / d{\rm ln}T$
for different approximations of the radiative losses in HD lines.}
\label{therm}
\end{figure}


\begin{thebibliography}{99}

\bibitem{abel97} T. Abel, P. Anninos, Y. Zhang and M.L. Norman, New Astr., {\bf
2}, 181 (1997)

\bibitem{bromm} V. Bromm, P. Coppi, R. Larson, R., Astrophys. J. {\bf 564}, 23
(2002)

\bibitem{t97} M. Tegmark, J. Silk, M.J. Rees, et al., Astrophys. J. {\bf 474}, 1
(1997)

\bibitem{grishukzeld} L. P.Grishchuk and Ya. B. Zel’dovich, Astron. Zh. 58,
472 (1981) [Sov. Astron. 25, 267 (1981)]

\bibitem{peebles1982} P.J.E. Peebles, Astrophys. J. {\bf 263}, L1, (1982)

\bibitem{kolb} E. Kolb and M. Turner, {\it The Early Universe}
(Addison-Wesley, Readwood City, 1990)

\bibitem{peebles} P. J. E. Peebles, {\it Principles of Physical Cosmology}
(Princeton Univ. Press, Princeton, 1993)

\bibitem{lin} C. Lin, L. Mestel, F. Shu,  Astrophys. J. {\bf 142}, 1431 (1965)

\bibitem{zeld} Ya. B. Zel’dovich, Astrofiz. 6, 119 (1970)

\bibitem{smith} J. Smith, Astrophys. J. {\bf 238}, 842 (1980)

\bibitem{coll} A. A. Suchkov, Yu. A. Shchekinov, and
M. A. . Edel’man, Astrofiz. 18, 629 (1982) [Astrophys.
18, 360 (1982)]

\bibitem{struck1} C. Struck-Marcel, Astrophys. J. {\bf 259}, 116 (1982)

\bibitem{struck2} C. Struck-Marcel, Astrophys. J. {\bf 259}, 127 (1982)

\bibitem{shapiro} P.R. Shapiro, H. Kang, Astrophys. J. {\bf 318}, 32 (1987)

\bibitem{yua1991} Yu. A. Shchekinov,  Astrophys. Space Sci. {\bf 175}, 57
(1991)

\bibitem{yamada} M. Yamada, R. Nishi, Astrophys. J. {\bf 505}, 148 (1998)

\bibitem{cen} R. Cen, Astrophys. J. {\bf 624}, 485 (2005) 

\bibitem{wmap} D. N. Spergel, L. Verde, H.V. Peiris et al, Astophys. J. Suppl. Ser.
{\bf 148}, 175 (2003)

\bibitem{lynden}  D. Lynden-Bell, Monthly Notices Roy. Astron. Soc. {\bf 136},
101 (1967)

\bibitem{miniati} F. Miniati, T. W. Jones, A. Ferrara and D. Ryu, Astrophys. J.
{\bf 491}, 216 (1997)

\bibitem{BL} R. Barkana, A., Loeb, Phys. Rept., {\bf 349}, 125 (2001)

\bibitem{gp} D. Galli, and F. Palla, Astron. Astropys. {\bf 335}, 403 (1998)

\bibitem{hm} D. Hollenbach and C.F. McKee, Astrophys. J. Suppl. Ser. {\bf 41}, 555
(1979)

\bibitem{flower} D. Flower, Monthly Notes Roy. Soc., {\bf 318}, 875 (2000)

\bibitem{puy} D. Puy, and M. Signore, NewA {\bf 3}, 247 (1998)

\bibitem{varsh} D. A. Varshalovich and V. K. Khersonskii, Pis’ma
Astron. Zh. 2, 574 (1976) [Sov. Astron. Lett. 2, 227
(1976)]

\bibitem{gp00} D. Galli, and F. Palla, Planet. Space Sci. {\bf 12-13}, 1197
(2002)

\bibitem{stone} M. Stone, Astrophys. J. {\bf 159}, 277 (1970)

\bibitem{gilden} D. Gilden, Astrophys. J. {\bf 279}, 335 (1984)

\bibitem{omukainishi} K. Omukai, R. Nishi, Astrophys. J. {\bf 508}, 141 (1998)

\bibitem{ripamonti} E. Ripamonti, F. Haardt, A. Ferrara, et al., Monthly Notices Roy. Astron. Soc. {\bf 334}, 401 (2002)

\bibitem{field} G.B. Field, Astrophys. J. {\bf 142}, 531 (1965)

\bibitem{yus78} Yu. A. Shchekinov, Astron. Zh. 55, 311 (1978) [Sov.Astron. 22, 182 (1978)]

\bibitem{omukaipalla} K. Omukai, F. Palla, Astrophys. J. {\bf 589}, 677 (2003)

\bibitem{kamaya02} H. Kamaya, J. Silk, Monthly Notices Roy. Astron. Soc. {\bf
332}, 251 (2002)

\bibitem{kamaya03} H. Kamaya, J. Silk, Monthly Notices Roy. Astron. Soc. {\bf
339}, 1256 (2003)

\bibitem{omukaikitayama}  K. Omukai, T. Kitayama, Astrophys.J. {\bf 599},
738-745 (2003)

\bibitem{omukaih2} H. Mizusawa, R. Nishi, and K. Omukai, Publ. Astron.
Soc. Jpn. (2005) (in press); astro-ph/0404333 (2004)

\end{thebibliography}
\end{document}